\documentclass{aa}
\usepackage{graphicx}
\usepackage{natbib}
\bibpunct{(}{)}{;}{a}{}{,}
\usepackage{txfonts}
\begin{document}
   \title{The CORALIE survey for southern extrasolar planets}

   \subtitle{XIV. HD~142022~b: a long-period planetary companion in a wide
   binary\thanks{Based on observations 
   collected at the ESO La Silla Observatory with the CORALIE echelle spectrograph mounted on the 
   Swiss telescope, and with the HARPS echelle spectrograph mounted on the ESO
   3.6-m telescope (programme ID 072.C-0488).}}

   \author{A.~Eggenberger\inst{1} \and M.~Mayor\inst{1} \and D.~Naef\inst{2} 
   \and F.~Pepe\inst{1} \and D.~Queloz\inst{1} \and N.C.~Santos\inst{1,3} 
   \and S.~Udry\inst{1} \and C.~Lovis\inst{1}
          }

   \offprints{Anne Eggenberger, \email{Anne.Eggenberger@obs.unige.ch}}

   \institute{Observatoire de Gen\`eve, 51 ch. des Maillettes, 
             1290 Sauverny, Switzerland
         \and
             European Southern Observatory, Alonso de Cordova 3107, Casilla
	     19001, Santiago 19, Chile
          \and
             Centro de Astronomia e Astrof\'{\i}sica da Universidade de Lisboa, 
	     Observat\'{o}rio Astron\'{o}mico de Lisboa, Tapada da Ajuda,
	     1349-018 Lisboa, Portugal  
	     }

   \date{Received / Accepted}

   \abstract{We report precise Doppler measurements of HD~142022 obtained 
   during the past six years with the CORALIE echelle spectrograph at La Silla 
   Observatory together with a few additional observations made recently with
   the HARPS echelle spectrograph. Our radial velocities reveal evidence of a planetary companion 
   with an orbital period $P=1928^{+53}_{-39}$~days, an eccentricity 
   $e=0.53^{+0.23}_{-0.18}$, 
   and a velocity semiamplitude $K=92^{+102}_{-29}$~m\,s$^{-1}$. The inferred 
   companion minimum mass is $M_2\sin{i}=5.1^{+2.6}_{-1.5}$~$\mathrm{M}_{\rm Jup}$ 
   and the semimajor axis $a=3.03\pm0.05$~AU. 
   Only one full orbital revolution has been monitored yet, and the 
   periastron passage could not be observed since the star was too low on the
   horizon. The eccentricity and velocity semiamplitude remain therefore quite 
   uncertain and the orbital solution is preliminary.  
   HD~142022 is a chromospherically inactive K0 dwarf, metal rich relative to
   the Sun, and is the primary component of a wide binary. HD~142022~b 
   is thus a new "planet in binary" candidate, and its high eccentricity might 
   be due to secular interactions with the distant stellar companion.

   }

   \titlerunning{A long-period planetary companion in a wide
   binary}
   
   \maketitle
%

\section{Introduction}

The CORALIE survey for southern extrasolar planets has been going on since June
1998. This high-precision radial-velocity programme makes use of the 
CORALIE fiber-fed echelle spectrograph mounted on the 1.2-m Euler Swiss 
telescope at La Silla Observatory (ESO, Chile). The sample of stars monitored 
for the extrasolar planet search is made of 1650 nearby G and K dwarfs 
selected according to distance in order to have a well-defined 
volume-limited set of stars \citep{Udry00}. The CORALIE sample can thus be used 
to address various aspects of the statistics of extrasolar planets.

Stellar spectra taken with CORALIE are reduced online. Radial velocities are
computed by cross-correlating the measured stellar spectra with a numerical 
mask, whose nonzero zones correspond to the theoretical positions and widths of
stellar absorption lines at zero velocity. The resulting cross-correlation 
function (CCF) therefore represents a flux weighted "mean" profile 
of the stellar absorption lines transmitted by the mask. The radial velocity of 
the star corresponds to the minimum of the CCF, which is determined by fitting
the CCF with a Gaussian function. 
The initial long-term velocity precision of CORALIE was about $7$~m\,s$^{-1}$ 
\citep{Queloz00}, but since 2001 the instrumental accuracy combined 
with the simultaneous ThAr-reference technique is better than $3$~m\,s$^{-1}$ 
\citep{Pepe02}. This implies that for many targets the precision is now 
limited by photon noise or stellar jitter.

After seven years of activity, the CORALIE survey has proven to be very
successful with the detection of a significant fraction of the known extrasolar 
planetary candidates (see the previous papers in this series). 
As the survey duration increases, new planetary
candidates with orbital periods of several years can be unveiled and their orbit
characterized. This is the case of the companion orbiting HD~142022:
with a period longer than 5 years it has just completed one orbital revolution 
since our first observation in 1999. This growing period-interval coverage is
very important with regard to formation and migration models since 
observational constraints are still very weak for periods of a few
years.

HARPS is ESO's High-Accuracy Radial-Velocity Planet Searcher
\citep{Pepe02b,Pepe04,Mayor03}, a
fiber-fed high-resolution echelle spectrograph mounted on the 3.6-m telescope at
La Silla Observatory (Chile). The efficiency and extraordinary 
instrumental stability of HARPS combined with a powerful data reduction pipeline
provides us with very high precision radial-velocity measurements, allowing the
detection of companions of a few Earth masses around solar-type stars
\citep{Santos04}. Benefiting from this unprecedented precision, 
a part of the HARPS Consortium Guaranteed-Time-Observations programme is 
devoted to the study of 
extrasolar planets in a continuation of the CORALIE survey, to allow 
a better characterization of long-period planets and multiple planetary systems.
HD~142022 is part of this programme, and the few HARPS measurements obtained so 
far already contribute to improve the orbital solution based on CORALIE data.

The stellar properties of HD~142022 are summarized in Sect. \ref{sect2}. 
Section \ref{sect3} presents our radial-velocity data for HD~142022 and the 
inferred orbital solution of its newly detected companion. These results are 
discussed in Sect. \ref{sect4}, showing that the planetary interpretation is 
the best explanation for the observed velocity variation. Our 
conclusions are drawn in Sect. \ref{sect5}.

\section{Stellar characteristics of HD~142022}
\label{sect2}

\object{HD~142022} (HIP~79242, Gl\,606.1\,A) is a bright K0 dwarf in the 
Octans constellation. 
The astrometric parallax from the Hipparcos catalogue, 
$\pi = 27.88 \pm 0.68$~mas (ESA 1997), sets the star at a
distance of $36$~pc from the Sun. With an apparent magnitude $V=7.70$ 
(ESA 1997) this implies an absolute magnitude of $M_{\rm V} = 4.93$. 
According to the Hipparcos catalogue the color index for HD~142022 is $B-V=0.790$. 
Using a bolometric correction $BC=-0.192$ \citep{Flower96} and the solar absolute
magnitude $M^{\rm bol}_{\rm \odot}=4.746$ \citep{Lejeune98} we thus obtain 
a luminosity $L=1.01$~$\mathrm{L}_{\rm \odot}$. 
The stellar parameters for HD~142022 are summarized in Table \ref{tab1}.

\begin{table}
\caption{Observed and inferred stellar parameters for HD~142022 (see text for
references).}
\begin{center}
\begin{tabular}{llc}
\hline\hline
Parameter & Unit & Value \\
\hline
Spectral Type               &                  & K0\\
$V$                         & (mag)            & 7.70\\
$B-V$                       & (mag)            & 0.790\\
$\pi$                       & (mas)            & 27.88$\pm$0.68\\
$M_{V}$               & (mag)            & 4.93\\
$T_{\rm eff}$               & (K)              & 5499$\pm$27\\
$\log{g}$                   & (cgs)            & 4.36$\pm$0.04\\
$[{\rm Fe}/{\rm H}]$        & (dex)            & 0.19$\pm$0.04\\
$L$                         & ($\mathrm{L}_{\odot}$)     & 1.01\\
$M_{\star}$                  & ($\mathrm{M}_{\odot}$)     & 0.99\\
$\upsilon\sin{i}$               & (km\,s$^{-1}$)   & 1.20\\
$\log(R^{\prime}_{\rm HK})$    &                  & $-4.97$\\
\hline
\end{tabular}
\end{center}
\label{tab1}
\end{table}

A detailed spectroscopic analysis of HD~142022 was performed using 
our HARPS spectra in order to obtain accurate atmospheric parameters 
(see \cite{Santos05} for further details). 
This gave the following values: an effective temperature 
$T_{\rm eff} = 5499\pm27$~K, a surface gravity $\log{g} = 4.36\pm0.04$, 
and a metallicity $[{\rm Fe}/{\rm H}] = 0.19\pm0.04$. 
Using these parameters and the 
Geneva stellar evolution code \citep{Meynet00} we deduce a mass 
$M_{\rm \star} = 0.99$~$\mathrm{M}_{\rm \odot}$.
According to evolutionary models, HD~142022 is an old main-sequence star, in
agreement with the K0V spectral type quoted in the Hipparcos catalogue.

The cross-correlation function can be used 
to derive stellar quantities affecting line profiles such as the projected rotational 
velocity. From the CORALIE spectra we derive $\upsilon\sin{i}=1.20$~km\,s$^{-1}$ 
\citep{Santos02}. Combining this result to the stellar radius given by the best 
evolutionary model ($R=1.15$~$R_{\rm \odot}$) we obtain an upper limit of 48~days for the 
rotational period. 

\begin{figure}
\resizebox{\hsize}{!}{\includegraphics[angle=-90]{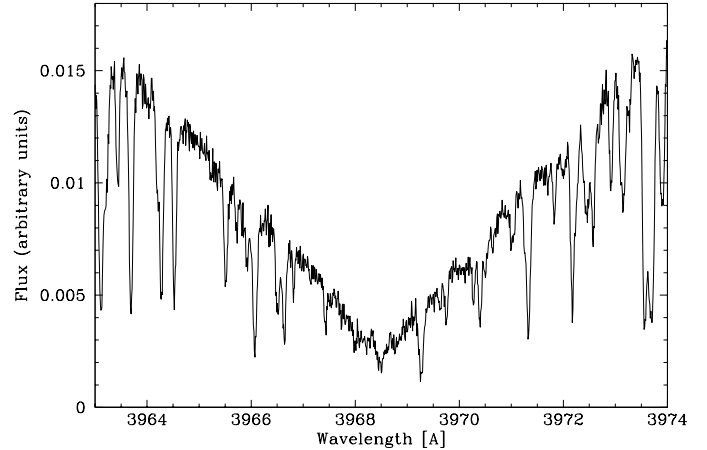}}
\caption{\ion{Ca}{ii} H ($\lambda=3968.47\,\AA$) absorption line region of the summed
HARPS spectra for HD~142022.}
\label{CaIIH_spectrum}
\end{figure}

From the HARPS spectra we can compute the $\log(R^{\prime}_{\rm HK})$ activity index 
measuring the chromospheric emission flux in the \ion{Ca}{ii} H and K lines. This index is a 
useful estimator of the radial-velocity jitter that can be expected from intrinsic 
stellar variability. Figure \ref{CaIIH_spectrum} shows the \ion{Ca}{ii} H absorption
line region for HD~142022. No emission peak is visible at the center of the
absorption line, indicating a rather low chromospheric activity. This is corroborated by 
the  $\log(R^{\prime}_{\rm HK})$ value of $-4.97$, typical of inactive stars. 

HD~142022 is part of a wide binary. Its companion, \object{LTT~6384} 
(Gl~606.1B), is a
late-K star about 22 arcseconds away and with an apparent magnitude $V=11.2$. 
The two stars are listed in the NLTT and LDS catalogs \citep{Luyten40,Luyten79} 
indicating very similar proper motions. They were also observed by the Hipparcos 
satellite, but the proper motion of LTT~6384 could not be determined. 
The apparent separation of the pair, nevertheless, remained close to 22
arcseconds from 1920 to 2000 which, given the proper motion of HD~142022 
($\mu_{\alpha}\cos{\delta}=-337.59\pm0.60$~mas\,yr$^{-1}$, 
$\mu_{\delta}=-31.15\pm0.72$~mas\,yr$^{-1}$), 
is an indication that the pair is indeed a bound system. This conclusion is 
strengthened by the fact that the CORAVEL radial velocities of the two stars are 
identical within uncertainties \citep{Nordstroem04}. Using the positions given
by the Tycho-2 catalogue and its supplement-2 (ESA 1997), we obtain a projected
binary separation of 820~AU. This translates into an estimated binary semimajor 
axis of 1033~AU, using the relation $a/r=1.26$\footnote{Strictly speaking, the 
relation used here to translate the projected separation of a wide binary into 
a semimajor axis is valid only statistically. It can thus be highly inaccurate 
for an individual system.} \citep{Fischer92}.

In the Hipparcos catalogue, HD~142022 is classified as an unsolved variable and is
suspected of being a non-single star. Indeed, a Lomb-Scargle periodogram of the 
Hipparcos photometry shows no clear signal standing out, but some extra power 
spread over many frequencies, especially at short period (few days). Performing 
a detailed study of the Hipparcos photometry for HD~142022 is beyond the 
scope of this paper, and is not fundamental since the periods involved are much 
shorter than that of the signal detected in radial velocity. We will nonetheless 
briefly come back to this issue in Sect. \ref{sect4}, because the potential 
non-single status of the star may be a concern.

\section{Radial velocities and orbital solution}
\label{sect3}

\begin{figure}
\resizebox{\hsize}{!}{\includegraphics{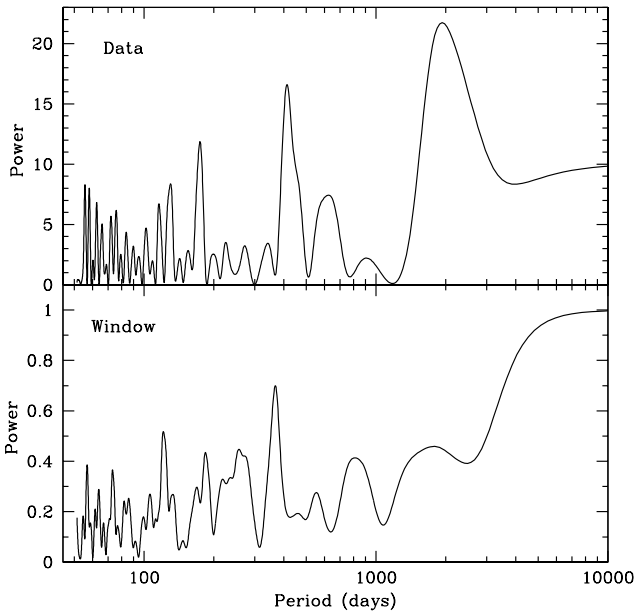}}
\caption{Lomb-Scargle periodogram of the CORALIE radial velocities for 
HD~142022 (top) and the corresponding window function (bottom). The power is a 
measure of the statistical significance of the signal, not of its true 
amplitude.}
\label{p_coralie}
\end{figure}

HD~142022 has been observed with CORALIE at La Silla 
Obervatory since July 1999. Altogether, 70 radial-velocity measurements with a 
typical signal-to-noise ratio of 25 (per pixel at 550~nm) and a mean 
measurement uncertainty (including photon noise and calibration errors) of 
9.7~m\,s$^{-1}$ 
were gathered. HD~142022 is also part of the HARPS high-precision 
radial-velocity 
programme \citep{Lovis05} and, as such, was observed 6 times between November 
2004 and May 2005. 
These observations have a typical signal-to-noise ratio of 90 and a mean measurement 
uncertainty of 1~m\,s$^{-1}$.  

\begin{figure}
\resizebox{\hsize}{!}{\includegraphics{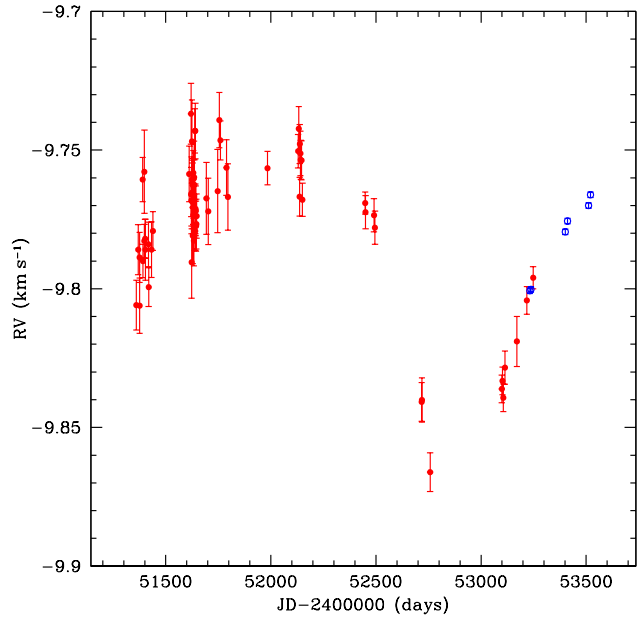}}
\caption{CORALIE and HARPS radial velocities for HD~142022. CORALIE 
velocities are shown as dots, HARPS data are plotted as circles. As
each instrument has its own zero point, a velocity offset of $-8.8$~m\,s$^{-1}$ 
has been added to the HARPS data (see text for further details). Error bars
for HARPS data (1~m\,s$^{-1}$) are the same size as the symbols.}
\label{vr}
\end{figure}

\begin{figure}
\resizebox{\hsize}{!}{\includegraphics{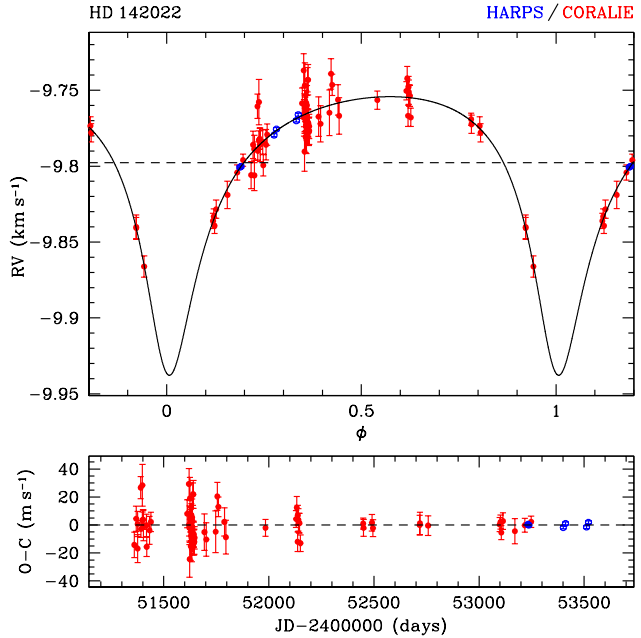}}
\caption{Phased CORALIE and HARPS radial velocities for HD~142022 (top). 
CORALIE observations are shown as dots, HARPS data are plotted as 
circles. The solid line is the best-fit orbital solution. Residuals to the
Keplerian fit are also displayed (bottom). Error bars
for HARPS data are the same size as the symbols.}
\label{kep_phas}
\end{figure}

The root-mean-square (rms) of the CORALIE radial velocities is 26.3 m\,s$^{-1}$, indicating some 
type of variability. The Lomb-Scargle periodogram of these velocities is shown 
in Figure \ref{p_coralie}. The highest peak corresponds to a period of 1926 
days, which is clearly visible in the plot of our radial velocities as a 
function of time (Fig. \ref{vr}). Using the expressions given in
\citet{Scargle82}, the false alarm probability for this signal is 
close to $10^{-8}$. This low value was confirmed using Monte Carlo simulations, 
in which data sets of noise only were generated with velocities drawn at random 
from the residuals around the mean. None of the $10^7$ simulated data set 
exhibited a maximum periodogram power exceeding the observed value, yielding a 
false alarm probability $<$$10^{-7}$. 
Figure \ref{kep_phas} shows the CORALIE and HARPS radial velocities phased with 
a similar period and the corresponding best-fit Keplerian model. The resulting 
orbital parameters are $P=1928$~days, 
$e=0.53$, $K=92$~m\,s$^{-1}$, implying a minimum mass 
$M_2\sin{i}=5.1$~$\mathrm{M}_{\rm Jup}$ orbiting with a semimajor axis 
$a=3.03$~AU. The orbital elements for HD~142022 are listed in Table \ref{tab3}. 
Since each data set has its own velocity zero point, the velocity offset 
between the two instruments (HARPS and CORALIE) is an additional free parameter
in the fitting process. Note that the first two HARPS measurements are
contemporary with the last CORALIE observations (Fig. \ref{vr}), and the 
curvature seen in the HARPS data fits perfectly well the CORALIE velocities 
taken during the previous revolution (Fig. \ref{kep_phas}). The velocity offset 
is therefore well constrained though the number of HARPS measurements is small. 

As can be seen in Fig. \ref{vr}, our data span only one orbital period 
and the phase coverage is very poor near periastron since the star was too 
low on the horizon to be observed at that time. This is why 
the orbital eccentricity and the velocity semiamplitude are poorly 
constrained. The orbital solution is thus 
preliminary and additional measurements taken during the next revolutions 
will be needed to obtain more accurate orbital parameters. It should be noted 
that for such an eccentric orbit the semiamplitude $K$ is strongly correlated 
with the eccentricity. As our present analysis is more likely to have 
overestimated the eccentricity, the semiamplitude $K$ might in fact be 
smaller, implying a smaller minimum mass for the companion. The companion
minimum mass is thus very likely to be in the planetary regime, whatever the
exact eccentricity.

The uncertainties in the orbital parameters were determined by applying the
fitting technique repeatedly to many sets of simulated data. 
1000 simulated data sets were thus constructed by adding a
residual value (drawn at random from the residuals) to the best-fit 
velocity corresponding to each observing time. For each realization, the 
best-fit Keplerian orbit was determined. This has been done by using the 
period obtained from the Lomb-Scargle periodogram as an initial guess for the 
Keplerian fit. We then used a local minimization algorithm to find the best 
fit, trying several initial starting values for $T$, $\omega$ and $K$. 
The quoted uncertainties correspond to the $1\sigma$ confidence interval of 
the resulting set of values for each orbital parameter.

The rms to the Keplerian fit is 10.8~m\,s$^{-1}$ for the CORALIE data and 
1.4~m\,s$^{-1}$ for the HARPS measurements, yielding a reduced $\chi^2$ of 1.5. 
The Keplerian model thus adequately explains the radial-velocity variation, 
though for both instruments the rms is slightly larger than the mean 
measurement uncertainties. A periodogram of the velocity residuals 
after subtracting off the best-fit Keplerian shows no signal with significant 
power anymore. Furthermore, no changes in stellar line profiles 
(as quantified by the CCF bisector span, 
see \citet{Queloz01}) are seen in our data.

\section{Discussion}
\label{sect4}

So far, it has been implicitly assumed that the radial-velocity variation 
observed for HD~142022 stems from the presence of a planetary companion. 
But it is well known that other phenomena can induce 
a periodic variation of the radial velocity of a star. Activity related
phenomena such as spots, plages or inhomogeneous convection are such 
candidates. They can however not be put forward to explain the  
signal observed for HD~142022 since the period is 1928~days, much too long to 
be related in any way to the rotational period of the star.  
Stellar pulsations may also cause radial-velocity variations, but no known 
mechanism could be invoked to sustain large-amplitude long-period oscillations in
a K0 main-sequence star. Stellar pulsations can therefore be ruled out as well.

The presence of a second and varying faint stellar spectrum superimposed on 
the target spectrum can also induce spurious radial-velocity variations 
mimicking a planetary-type signature (see the seminal case of
HD~41004, \citet{Santos02,Zucker03}). 
In such a case, the target is the unrecognized
component of a multiple stellar system. Given the fact that HD~142022 was 
suspected of being non-single on the basis of Hipparcos data, this is a 
possibility to take seriously into consideration.
The evolution of the cross-correlation 
bisector span as a function of radial velocity can be a useful 
tool in disentangling planetary signatures from spurious radial-velocity 
signals \citep{Queloz01,Santos02}. 
For a signal of planetary origin the bisector span is
constant whatever the value of the radial velocity, whereas for blended systems 
it is often correlated with the measured radial velocity 
(see for example Fig. 3 of \citet{Santos02}). No such correlation is visible  
for HD~142022. We have also searched for the presence of a second spectrum in 
our CORALIE spectra using multi-order TODCOR, a two-dimensional 
cross-correlation algorithm \citep{Mazeh94,Zucker03}, 
but did not find anything convincing. 
These negative results do not allow us to formally discard the blend scenario, 
but they render it an unlikely possibility. 
To sum up, the 1928-day signal observed in our radial velocities for
HD~142022 is most likely to be caused by the gravitational perturbation of a
planet orbiting the star.

\begin{table}
\caption{Orbital parameters for HD~142022~b.}
\begin{center}
\begin{tabular}{llc}
\hline\hline
Parameter & Unit & Value \\
\hline
$P$                         & (days)            & $1928^{+53}_{-39}$\\
$T$                       & (JD-2400000)      & $50941^{+60}_{-91}$\\
$e$                         &                   & $0.53^{+0.23}_{-0.18}$\\
$\gamma$                    & (m\,s$^{-1}$)     & $-9.798^{+0.007}_{-0.010}$\\
$\omega$                    & (deg)             & $170^{+8}_{-10}$\\
$K$                         & (m\,s$^{-1}$)     & $92^{+102}_{-29}$\\
$M_2\sin{i}$              & ($\mathrm{M}_{\rm Jup}$)   & $5.1^{+2.6}_{-1.5}$\\
$a$                         & (AU)              & $3.03^{+0.05}_{-0.05}$\\
Velocity offset (HARPS)     & (m\,s$^{-1}$)     & $-8.8^{+2.5}_{-2.5}$\\
\cline{1-3}
$N_{\rm meas}$ (CORALIE+HARPS)&                   & 70+6\\
rms (CORALIE)               & (m\,s$^{-1}$)     & 10.8\\
rms (HARPS)                 & (m\,s$^{-1}$)     & 1.4\\
\hline
\end{tabular}
\end{center}
\label{tab3}
\end{table}

The Keplerian fit to the radial velocities of HD~142022 implies that the
stellar orbit has a semimajor axis $a_1\sin{i} = 0.015$~AU. 
Given the stellar parallax, this translates into an angular semimajor 
axis $\alpha_1\sin{i} = 0.41$
mas. If the true mass of HD~142022~b were much larger than its minimum mass of
5.1~$\mathrm{M}_{\rm Jup}$, the stellar wobble might be detected in the 
Hipparcos astrometry. This potential wobble may, however, be partially absorbed 
into the solution for proper motion and parallax since the orbital period is 
longer than the 2.7-year duration of the Hipparcos measurements. 
We searched for an astrometric wobble in the Hipparcos data for HD~142022, but 
did not find anything significant.

HD~142022~b is one of the planets with the longest orbital period found in 
a wide binary so far, but its separation of 3~AU is still very small compared to the 
estimated binary semimajor axis of 1033~AU. HD~142022~b thus orbits well 
inside the stability zone, whatever the exact orbital parameters of the binary
\citep{Holman99}. The presence of a distant stellar companion may
nonetheless cause significant secular perturbations to the planetary orbit. In
particular, if the planetary orbital plane is inclined relative to the binary
plane, the planet can undergo large-amplitude eccentricity oscillations due 
to the so-called Kozai mechanism (\citet{Kozai62}; see also
\citet{Holman97,Innanen97,Mazeh97}). The Kozai mechanism is effective at very 
long range, but its oscillations may be suppressed by other competing sources 
of orbital perturbations, such as general relativity 
effects or perturbations resulting from the presence of an additional companion 
in the system. Regarding HD~142022, we have estimated the ratio $P_{\rm
Kozai}/P_{\rm GR}$ using equations 3 and 4 of \citet{Holman97} with the values
$e_{\rm b}=1/\sqrt{2}$ and $M_{\rm s}=0.6$~$\mathrm{M}_{\odot}$ for the binary 
eccentricity and secondary component mass, respectively. This yields 
$P_{\rm Kozai} = 1.25\,10^{8}$ years and 
$P_{\rm Kozai}/P_{\rm GR} = 0.35$, indicating that Kozai oscillations could 
take place is this system, since their period is shorter than the
apsidal period due to relativistic effects. Although not well constrained, the
eccentricity of HD~142022 is clearly quite high, and such a high eccentricity 
is not surprising if the system undergoes Kozai oscillations.

\section{Conclusion}
\label{sect5}

We report a 1928-day radial-velocity variation of the K0 dwarf HD~142022
with a velocity semiamplitude of 92~m\,s$^{-1}$. From the
absence of correlation between stellar activity indicators and radial
velocities, and from the lack of significant spectral line asymmetry
variations, the
presence of a planetary companion on a Keplerian orbit best explains our data. 
The Keplerian solution results in a $M_2\sin{i}=5.1$~$\mathrm{M}_{\rm Jup}$
companion orbiting HD~142022 with a semimajor axis $a=3.03$~AU and an 
eccentricity $e=0.53$. Although HD~142022~b orbits the primary component of a
wide binary, its characteristics, including minimum 
mass and orbital eccentricity, are typical of the long-period planets found 
so far around G and K dwarfs. 

One of the most surprising properties of extrasolar planets revealed by 
ongoing radial-velocity surveys is their high orbital eccentricities, which
challenge our current theoretical paradigm for planet formation. 
Several mechanisms have thus been proposed to account for eccentric planetary
orbits. One of them is the Kozai mechanism, a secular interaction between a 
planet and a wide binary companion in a hierarchical triple system with 
high relative inclination. Although the Kozai mechanism 
can be put forward to explain the high eccentricity of a few planetary 
companions: 16~Cyg~Bb \citep{Holman97,Mazeh97}, HD~80606~b \citep{Wu03} 
and possibly HD~142022~b, it seems impossible to explain the observed 
eccentricity distribution of extrasolar planets solely by invoking the 
presence of binary companions \citep{Takeda05}. According to \cite{Takeda05}, 
Kozai-type perturbations could nonetheless play an important
role in shaping the eccentricity distribution of extrasolar planets,
especially at the high end. In this regard, ongoing programmes aiming at
searching for new (faint) companions to stars with known planetary systems, or
aiming at estimating the frequency of planets in binary systems should soon 
bring new observational material, and enable us to refine our present knowledge.

\begin{acknowledgements}
      We thank S. Zucker for his help in searching for an astrometric 
      wobble in the Hipparcos data. 
      We also thank R. Behrend, M. Burnet, B. Confino, C. Moutou, B. Pernier, 
      C. Perrier, D. S\'egransan and D. Sosnowska for having carried out some 
      of the observations of HD~142022. 
      We are grateful to the staff from the Geneva Observatory, in particular to
      L. Weber, for maintaining the 1.2-m Euler Swiss telescope and the 
      CORALIE echelle spectrograph at La Silla, and for technical support during
      observations. 
      We thank our Israeli colleagues, T. Mazeh, B. Markus and S. Zucker, for 
      providing us with a version of their multi-order TODCOR code, and for 
      helping us running it on CORALIE spectra. 
      We thank the Swiss National Research Foundation (FNRS) and the Geneva
      University for their continuous support to our planet search programmes. 
      Support from Funda\c{c}\~{a}o para a Ci\^encia e a Tecnologia (Portugal)
      to N.C. Santos in the form of a scholarship (reference
      SFRH/BPD/8116/2002) and a grant (reference POCI/CTE-AST/56453/2004) is
      gratefully acknowledged. 
      This research has made use of the VizieR catalogue access tool operated at
      CDS, France.
\end{acknowledgements}


\bibliographystyle{aa} 
\bibliography{3720bib}

\end{document}